\documentclass{aastex62}
\usepackage{amsmath}
\usepackage{amssymb}
\usepackage[normalem]{ulem}
\usepackage{appendix}
\usepackage{graphicx}
\usepackage{natbib}
\def\be{\begin{equation}}
\def\ee{\end{equation}}
\def\bea{\begin{eqnarray}}
\def\eea{\end{eqnarray}}

\def\rsun{R$_\odot$}

\begin{document}

\shorttitle{Changes in the solar rotation}
\shortauthors{Basu \& Antia}

\title{Changes in the solar rotation over two solar cycles}

\author[0000-0002-6163-3472]{Sarbani Basu}
\affiliation{Department of Astronomy, Yale University, New Haven, CT, 06520, USA}
\email{sarbani.basu@yale.edu}
\author[0000-0001-7549-9684]{H. M. Antia}
\affiliation{Tata Institute of Fundamental Research, Homi Bhabha Road, Mumbai 400005, India}

\begin{abstract}
We use helioseismic data from ground and space-based instruments to analyze how
solar rotation has changed since the beginning of solar Cycle~23 with emphasis
on studying the differences between Cycles 23 and 24.  We find that the nature
of solar rotation is indeed different for the two cycles. While the changes in the
latitudinally independent component follows solar-cycle indices, some of the
other components have a more complicated behavior.  There is a substantial change
in the behavior of the solar zonal flows and their spatial gradients too. While
the zonal flows are in general weaker in Cycle-24 than  those in Cycle~23, there are clear
signs of the emergence of Cycle~25.  We have also investigated the properties
of the solar tachocline, in particular,  its position, width, and the change (or
jump) in the rotation rate across it.  We find significant temporal variation
in the change of the rotation rate across the tachocline. We also find that the
changes in solar Cycle~24 were very different from those of Cycle~23. We do not
find any statistically significant change in the position or the width of the
tachocline.

\end{abstract}

\keywords{Sun:helioseismology --- Sun:oscillations --- Sun:interior --- Sun:rotation}

\section{Introduction}
\label{sec:intro}

Helioseismic data allow us to determine changes that occur inside the Sun on
time scales of months and years and hence can be used to probe how the Sun
changes over solar activity cycles. Even before the availability of detailed
helioseismic data, \citet{howard1980} found that the rotation rate at the solar
surface varies with time and the pattern was referred to as torsional oscillations
or zonal flows. With the availability of continuous helioseismic data this
pattern was also detected in the subsurface rotation rate 
\citep{agkjs1997, schou1999,antiabasu2000,rachel2000,vorontsov2002}. 
The most prominent feature in these results is the migration of bands of
faster and slower than average rotation moving towards the equator at low latitudes.
\citet{antiabasu2001} found that at high latitudes these bands move towards the poles.
This pattern has been extensively studied over the solar Cycle~23 \citep[e.g.,][]{hma2008}
and it was thought that the pattern may extend to other cycles with
a period of about 11 years. However, later observations have revealed that there
are significant differences between the features observed in Cycles~23 and 24.
Now that we are close to the end of solar Cycle~24, in this work we use global helioseismic
data collected since 1995 to examine how solar dynamics has changed, in
particular, we study the differences in the internal dynamics of the Sun
during solar Cycles~23 and 24.

The minimum between solar Cycles 23 and 24 was deeper than any since the early
twentieth century. It was the quietest minimum recorded in the era of detailed
data: it had more sunspot-free days than any recorded in the space age, the
10.7-cm flux was the lowest ever recorded and the polar fields were very weak
too.  Cycle~24 that followed has been quite different from Cycle~23 and it has
been a much weaker cycle.  \citet{basuetal2012} reported that low-degree
helioseismic data indicated that Cycle~24 would be very different, and recently
\citet{howeetal2017} have shown that the difference in characteristics has
continued to date. An early examination of solar rotation during the minimum
just before Cycle~24 \citep{antiabasu2010, antiabasu2013} has revealed that the
solar rotation profile was different from that of the minimum before Cycle~23,
and studies show that the differences have continued \citep[see
e.g.,][]{howeetal2013, rudi2014, rachel2018, sasha2019}. We present the results
of an independent helioseismic study of solar dynamics using both ground-based
and space-based helioseismic data. Unlike most of earlier works that have
looked into differences between solar Cycles 23 and 24, we do not confine our
study to near-surface layers alone, but also study the tachocline which is
believed to be the seat of the solar dynamo \citep[see e.g.,][and references
therein]{gilman}. We use a mixture of helioseismic inversions and forward
modelling to do so.

The rest of the paper is organized as follows: in Section~\ref{sec:data} we
list the data used; Section~\ref{sec:rot} is devoted to the mean rotation rate.
We discuss zonal flows in Section~\ref{sec:zon}, the tachocline in
Section~\ref{sec:tach}, and finally in Section~\ref{sec:conc} we summarize our
findings.

\section{Data used}
\label{sec:data}

Helioseismic data consist of frequencies of solar oscillations $\nu_{nlm}$,
where $n$ is the radial order of the mode, or number of nodes in the interior,
$l$ the degree, i.e., number of nodes on the surface and $m$ the
azimuthal order, i.e., the number of nodes along the equator. It is more usual
to express the frequency $\nu_{nlm}$ as
\be
\nu_{nlm}=\nu_{nl}+\sum_{j=1}^{j_{\rm max}} c_j^{(n,l)}{\mathcal P}^{l}_j(m),
\label{eq:freq}
\ee
where the central frequency $\nu_{nl}$ depends on the structure of the Sun, and
the odd-order `splitting' coefficient $c_1$, $c_3$, $c_5$ etc., depend on the
internal rotation, and ${\mathcal P}^{l}_j$ are polynomials of degree $j$ in
$m$ \citep{ritzwoller1991}. Information about rotation, including the tachocline is coded in the
odd-order coefficients.

We use  helioseismic data from three sources: (I) The ground-based Global
Oscillation Network Group (GONG) \citep{hill1996}, (II) the Michelson Doppler Imager
(MDI) on board the Solar and Heliospheric Observatory spacecraft \citep{mdi}
and (III) the Helioseismic and Magnetic Imager (HMI) \citep{hmi}  on board the
Solar Dynamics Observatory.

The GONG data we use cover a period from May 5, 1995 to February 5, 2019. The data
are designated by GONG ``months'', each ``month'' being 36 days long. Solar oscillation
frequencies and splittings of sets starting Month~2 are obtained using 108 day
(i.e., 3 GONG months) time series. There is an overlap of 72 days between
different data sets, i.e., GONG Month~2 frequencies were obtained from data of
GONG Months~1, 2 and 3, those for Month~3 from GONG Months~2, 3 and 4, etc.
GONG data are available with solar $m$-dependent frequencies as well as
splitting coefficients of different kinds, we downloaded the frequencies
$\nu_{nlm}$ and fitted them to Eq.~(1) to obtain the splitting coefficients,
$c_j$ as defined by \citet{ritzwoller1991}. All data sets are publicly
available and can be obtained from the data archives at https://gong.nso.edu.

Data from the MDI cover the period from May 1, 1996 to April 24, 2011. Solar oscillation
frequencies and splittings for these data are obtained from 72 day time series
and the sets have no overlap in time. HMI started obtaining data on April 30,
2010. Like MDI, frequencies and splittings are obtained from non-overlapping
72-day time series. We use data obtained until March 13, 2019. MDI and HMI
splitting coefficients have a somewhat different definition from the Ritzwoller
\& Lavely splitting coefficients. To keep all data in the same form and to ensure that the
results can be compared to previous work and to data from GONG, we have converted the
splittings to the Ritzwoller \& Lavely form.  Data from MDI and HMI are
publicly available from the Joint Science Operations Center at Stanford
(https://jsoc.stanford.edu).
For the purpose of studying temporal variation in the rotation rate, or the zonal
flows, we have combined the MDI and HMI data. In the overlapping one year between
the two sets, we use HMI data. This assumes that there are no systematic differences
between the two sets \citep{larson2018}. However, as pointed out later we do find some
differences between the two, and this manifests in the zonal flows at high
latitudes.

We also use data on solar activity indices, in particular, the radio flux at
10.7~cm (\citealt{tapping2} and \citealt{tapping})\footnote{Available from
 http://www.spaceweather.gc.ca/solarflux/sx-en.php and 
https://omniweb.gsfc.nasa.gov/form/dx1.html} and the international sunspot 
number (SSN) from 
\citet{sidc}\footnote{Available from http://www.sidc.be/silso/datafiles}. 
We also use data on sunspot positions from Royal Observatory, Greenwich
database\footnote{Available
from http://solarscience.msfc.nasa.gov/greenwch.shtml} and  the Debrecen Photoheliographic 
Data\footnote{Available from http://fenyi.solarobs.csfk.mta.hu/DPD/} (see
\citealt{baranyi1} and \citealt{baranyi2}).

\section{Solar rotation}
\label{sec:rot}

\begin{figure}
\epsscale{0.90}
\plottwo{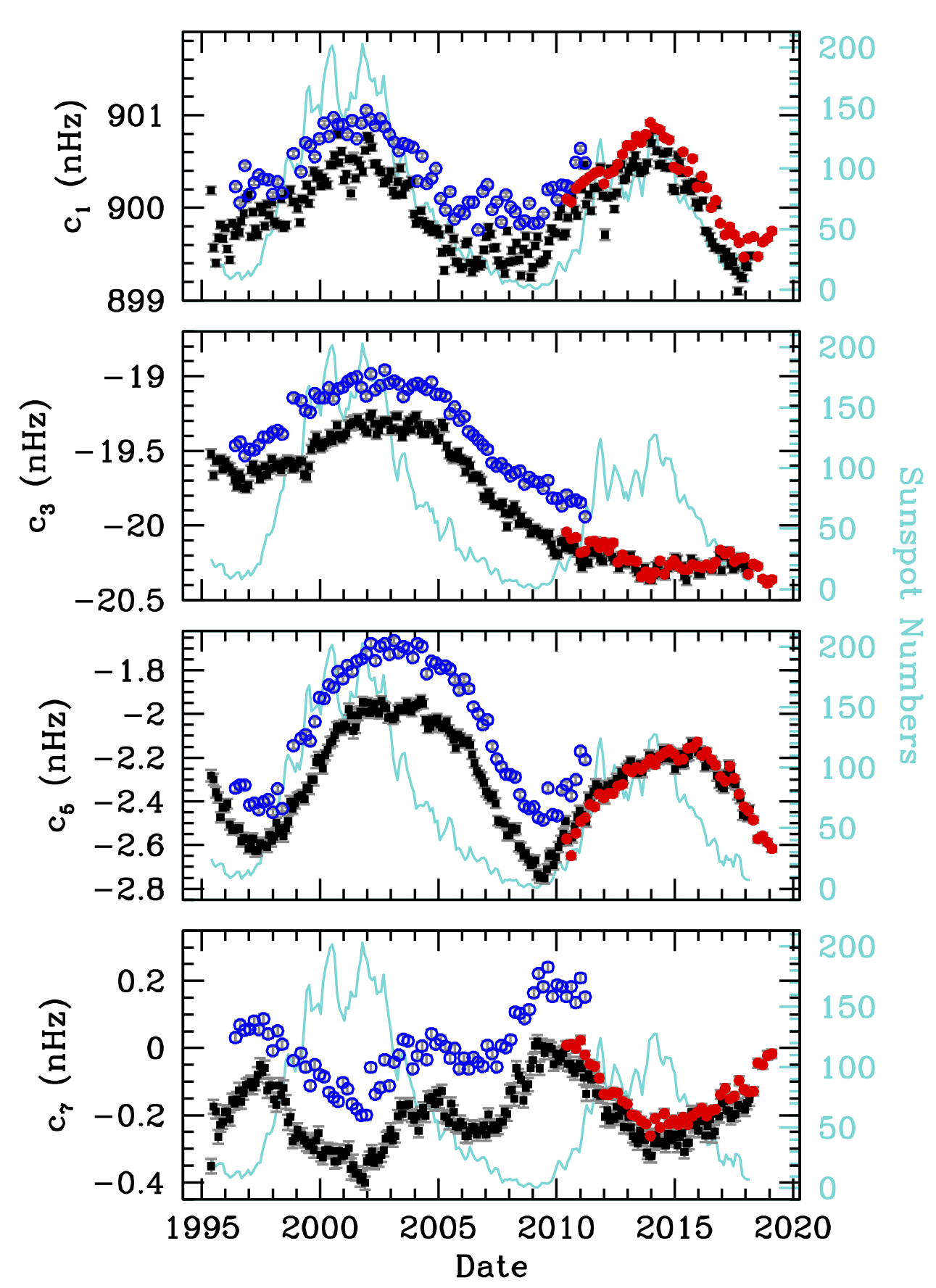}{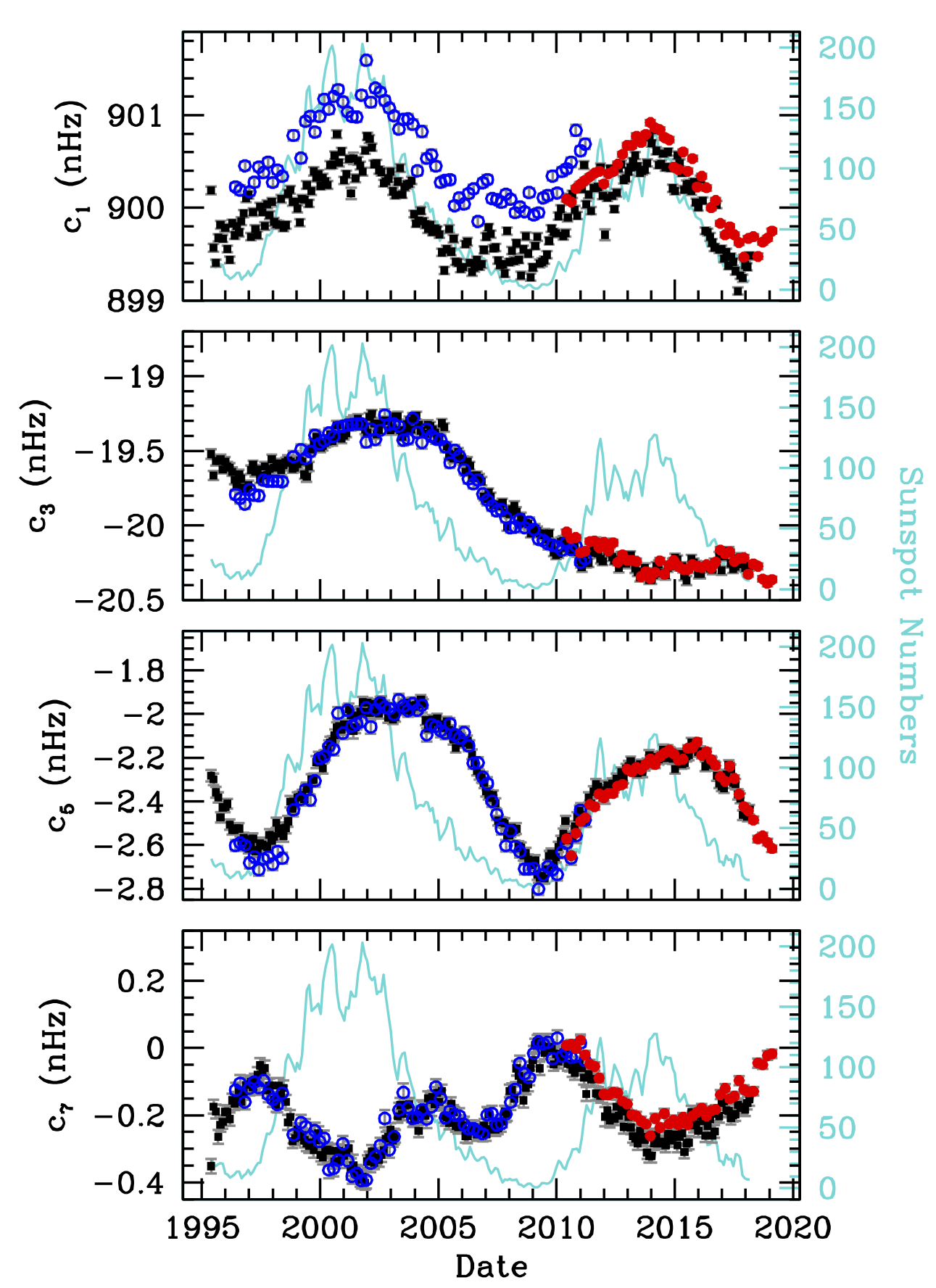}
\caption{ The change of splitting coefficients as a function of time.
The points represent averaged splitting coefficients (from top to
bottom: $c_1$, $c_3$, $c_5$ and $c_7$) for modes with lower turning points
between 0.95\rsun and 0.975\rsun. The black points are GONG data, blue MDI and
red HMI. The column on the left are average of all data, while in the
right-hand column we have restricted MDI data to modes with degree $l < 120$. In
all panels, the cyan lines show SSN values averaged over time intervals that
correspond to the GONG sets as the light blue line, with the values on the
right-hand axis.
}
\label{fig:coef}
\end{figure}

The solar rotation rate is obtained by inverting the odd-order splitting
coefficients \citep{schou1998}, however, the time-variations can be seen easily
in the splittings themselves. In the left-hand column of Fig.~\ref{fig:coef} we
show the first four odd-order splitting coefficients averaged over all modes that
have lower-turning points between 0.950\rsun\ and 0.975\rsun. A number of
features are clear immediately. First, there are clear systematic differences
between MDI and GONG data, much less so between HMI and GONG data. However, the
systematic differences are smaller when MDI data are restricted to modes
with $l < 120$, as can be seen in the right-hand panel of Fig.~\ref{fig:coef}.
The problem with high-degree MDI modes were reported earlier by \citet{ab2004} and
\citet{hma2008}, and it appears that reprocessing by \citet{larson} has not fully resolved
that.  
A part of the remaining difference between MDI and GONG, particularly for $c_1$ 
is because of the f modes in the MDI data.  Similarly, some of the differences between GONG and 
HMI are because GONG data are restricted to $l <= 150$, and because of f modes
that are not present in the GONG sets.

{ It is known that there are systematic differences between GONG and MDI
data for rotational splittings \citep{schou2002} which yields a slightly different
rotation profiles for the two data sets. This differences persist even if $l>120$ modes
are neglected. Most of these systematic differences are believed to be due to the
difference in processing pipeline \citep{schou2002} and are independent of time.
As a result, the zonal flows obtained by subtracting the temporal average are not
affected significantly by these differences. However, some differences do show
up in the gradients of zonal flows \citep{hma2008}, particularly in near
surface layers. These differences can be reduced if $l>120$ modes are neglected
as pointed out by \citet{hma2008}. All results shown in this work are obtained
by neglecting $l>120$ modes for MDI.}

\begin{deluxetable*}{lccccccc}
\tablecolumns{8}
\tablecaption{Correlation coefficient of splitting coefficients with 10.7 cm radio flux
\label{tab:corr}}
\tablehead{\colhead{Coefficient}& \colhead{Both}& \multicolumn{3}{c}{Cycle~23} &
\multicolumn{3}{c}{Cycle~24}\\
\colhead{} & \colhead{Cycles} & \colhead{Full} & \colhead{Ascending} & \colhead{Descending}
& \colhead{Full} & \colhead{Ascending} & \colhead{Descending}}
\startdata
$c_1$ & $\phantom{-}0.79$ & $\phantom{-}0.87$ & $\phantom{-}0.83$ & $\phantom{-}0.92$ & $\phantom{-}0.82$ & $\phantom{-}0.77$ & $\phantom{-}0.90$\\
$c_3$ & $\phantom{-}0.36$ & $\phantom{-}0.72$ & $\phantom{-}0.84$ & $\phantom{-}0.75$ & $-0.50$ & $-0.72$ & $-0.34$ \\
$c_5$ & $\phantom{-}0.64$ & $\phantom{-}0.63$ & $\phantom{-}0.91$ & $\phantom{-}0.71$ & $\phantom{-}0.64$ & $\phantom{-}0.88$ & $\phantom{-}0.67$ \\
$c_7$ & $-0.71$ & $-0.79$ & $-0.93$ & $-0.62$ & $-0.68$ & $-0.87$ & $-0.84$\\
\enddata
\end{deluxetable*}

The more interesting features as they relate to the Sun however, are the time
variations. For modes that sample this radius range,  the $c_1$ coefficient
appears to change with solar cycle and are largest at the maximum of the cycle;
 $c_5$ has a similar behavior. Coefficient $c_3$ however, does not
show such a dependence on the 11-year Schwabe cycle, and its time dependence,
positive correlation with solar activity indices in Cycle~23 and negative in
Cycle~24, makes it tempting to speculate whether it follows a cycle with a
longer time scale or shows a more complicated variation. The behavior of the coefficients during Cycle~25 should 
clarify this.
Coefficient $c_7$ has a  more complicated time variation;
in Cycle~24 the coefficient is anticorrelated with solar
activity, in Cycle~23 there is a hump-like feature around 2005; since
the feature 
is seen both by GONG and MDI it must be a feature of the Sun and not an
instrumental effect.  

A correlation analysis of the coefficients with the
10.7~cm flux (Table~\ref{tab:corr}) shows that for the modes shown in the figures, $c_1$ is positively
correlated with 
 the 10.7 cm flux in both cycles with the correlation coefficient being as high
as $0.9$ in the descending phases of the two cycles. Coefficient $c_5$ is also
positively correlated, but more so in the ascending phases of the two cycles,
Additionally, $c_5$ seems to show the
sharpest change as a new cycle begins --- at the start of both Cycle~23 and
Cycle~24, $c_5$ reversed its fall abruptly and started increasing; in contrast, the change in
$c_1$ was gentler.
{Extrapolating the curve for $c_5$ it appears that some time in early 2020
it will reach the level that was reached at the minimum around 2009. This could
be the time for next solar activity minimum though there is some ambiguity in
this criterion.}

Note that after solar maximum, there is a noticeable
time-lag between the fall of the radio flux and $c_5$, this is unlike the
case of $c_1$, which starts falling as soon as the radio flux does.
The coefficient $c_7$ shows a more complicated behavior during Cycle~23 with strong
negative correlation with the 10.7 cm flux during the ascending part of
the cycle, which becomes weaker during the descending part
of Cycle~23 possibly because of the feature around {2005}.
 Just like $c_5$, $c_7$ changes quite abruptly 
at the onset of the new cycle, with the splittings decreasing. The onset
of Cycle~25 should be able to confirm if the abrupt changes in $c_5$ and $c_7$
are persistent features of all solar cycles.

\begin{figure}
\epsscale{0.6}
\plotone{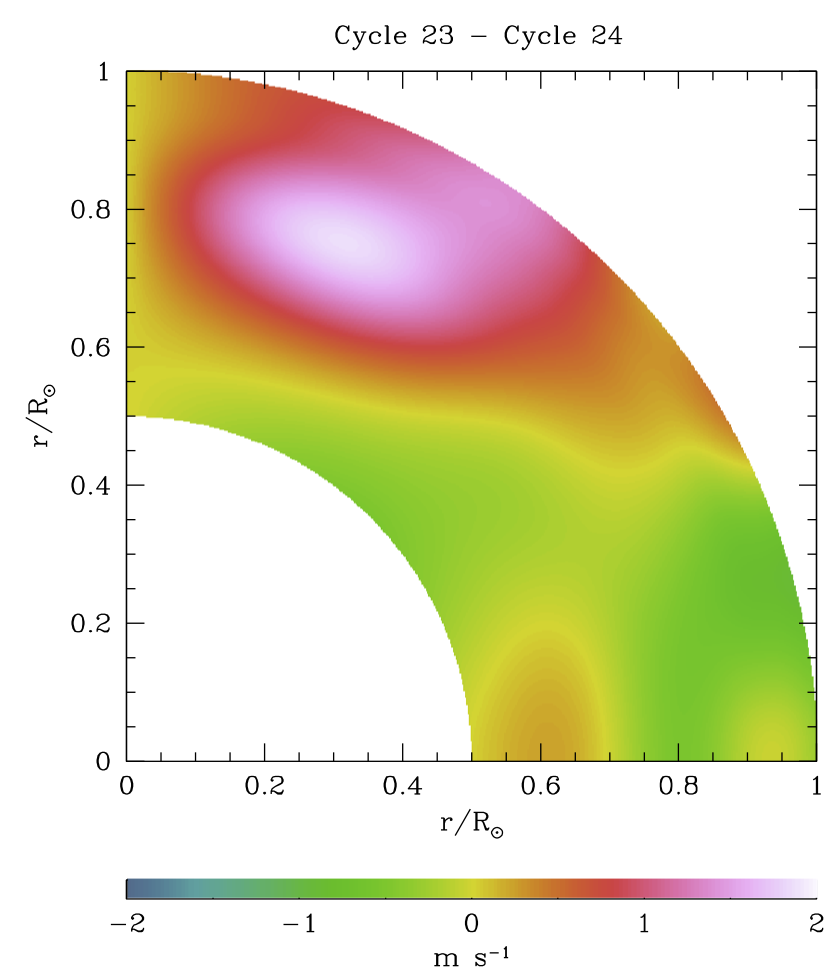}
\caption{ The difference between the average rotational velocities during
solar Cycle~23 and solar Cycle ~24. The differences are in the sense
(Cycle~23~$-$~Cycle~24).  Results are for GONG data.
}
\label{fig:gongdiff}
\end{figure}

The differences in the time-variations of the splitting coefficients imply that
the average rotation rate during Cycle~23 was different from that of Cycle~24,
and that the differences will be latitude-dependent, which is
indeed the case as shown in Fig.~\ref{fig:gongdiff}. As expected, there are
significant differences; in the  convection zone, Cycle~23 had lower rotation
velocities at the active latitudes but higher velocities at higher latitudes
compared with Cycle~24. It should be noted that the differences between the
average rotation rates are smaller than the difference in the rotation rate
between the minima before Cycle~23 and Cycle~24 \citep{antiabasu2010}, but
these results, as we shall see in the next section, have important implications
for solar zonal-flows.

\section{Zonal flows and their gradients}
\label{sec:zon}

\begin{figure}
\epsscale{0.75}
\plotone{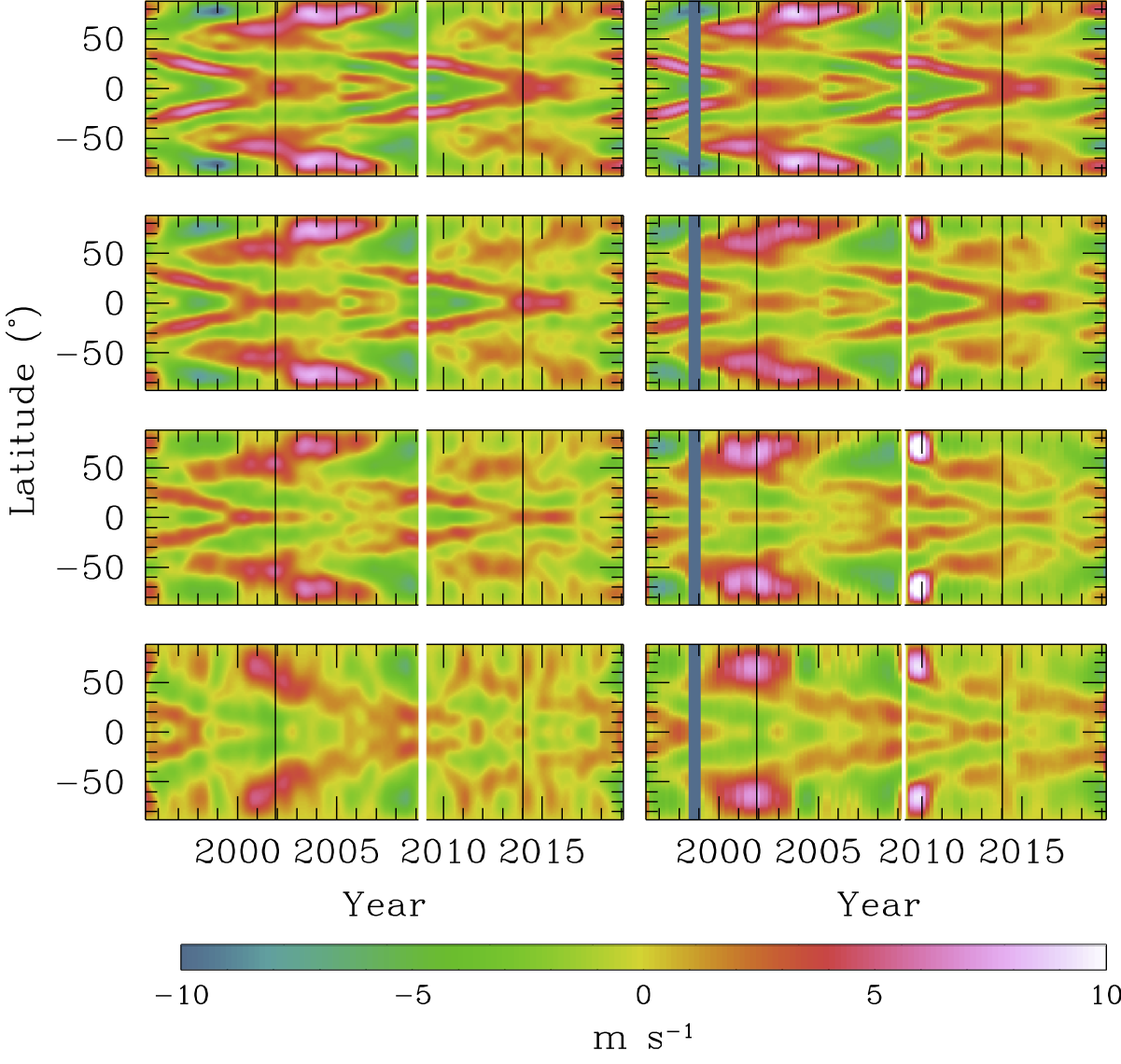}
\caption{Zonal flow velocities plotted as a function of time and latitude at,
from top to bottom, radii of 0.98, 0.95, 0.90 and 0.80\rsun. GONG results are 
shown in the left column, and MDI and HMI results in the right column, In both columns,
the thin white vertical patch demarcates Cycle~23 and 24. The {blue-green} vertical patch in
the right column covers the time when MDI data were not available because the
SOHO spacecraft was out of contact. The two black vertical lines in each panel mark the time of
maximum solar activity. The results have been smoothed over a year to
reduce the fluctuations that are present in the data sets.
The uncertainties in the velocities at low latitudes is about 0.2 m s$^{-1}$ at 0.98\rsun and increases to 0.4 m s$^{-1}$
at 0.8\rsun.
}
\label{fig:zonl}
\end{figure}

\subsection{The analysis}
\label{subsec:zonanal}

Zonal flows are east-west flows in the Sun with alternating bands of prograde 
and retrograde flows.
Surface observations showed that bands in the active latitudes migrate towards
the equator, while higher-latitude bands migrate towards the poles with time
\citep{howard1980,ulrich2001}. These flows are present below the surface too
(first seen in f modes, \citealt{agkjs1997}), and these too migrate to
different latitudes with time \citep{schou1999}.

Helioseismic studies of zonal flows define them to be residuals left at any epoch
when the average rotation velocity is subtracted out, i.e.,
\be
\delta v_\phi = v_\phi(r,\theta,t)- \langle v_\phi(r,\theta, t)\rangle,
\label{eq:zonal}
\ee
where $v_\phi(r,\theta, t)\equiv \Omega(r,\theta,t)r\cos\theta$ is the rotation
velocity at a given position and time in the Sun, and the angular brackets
represent the time-average. Here, $\theta$ is the latitude. As is clear from Eq.~\ref{eq:zonal}, the results
depend on the time over which the averaging is done and hence the
weaker features are not very robust.  If the average over all data 
are used, the higher rotation rate of Cycle~23 in
the high-latitude regions washes out some of the weaker features of the zonal
flow during Cycle~24, in particular, one cannot see the high-latitude poleward
branch of the zonal flow during Cycle~24 \citep{antiabasu2013, howe2013, rachel2018}, this had
earlier led to speculation that this might mean that Cycle~25 may be delayed
\citep{hill2011}.  Consequently, in this work we treat Cycles 23 and 24
separately.

The differences in the zonal flows can be made clearer by analyzing the spatial
gradients of the zonal flows. In keeping with \citet{hma2008}, we calculate the
gradients of $\delta\Omega$ rather than $\delta v_\phi$. The radial gradient is
simply $\delta\Omega_r\equiv \partial\delta\Omega/\partial r$, and the
latitudinal gradient is $\delta\Omega_\theta\equiv
(1/r)\partial\delta\Omega/\partial|\theta|$. However, in the results shown
later these gradients are multiplied by $\cos\theta$ to balance the latitudinal
dependence as in the case of zonal flows.

\subsection{Results}
\label{subsec:zonres}

\begin{figure}
\epsscale{0.75}
\plotone{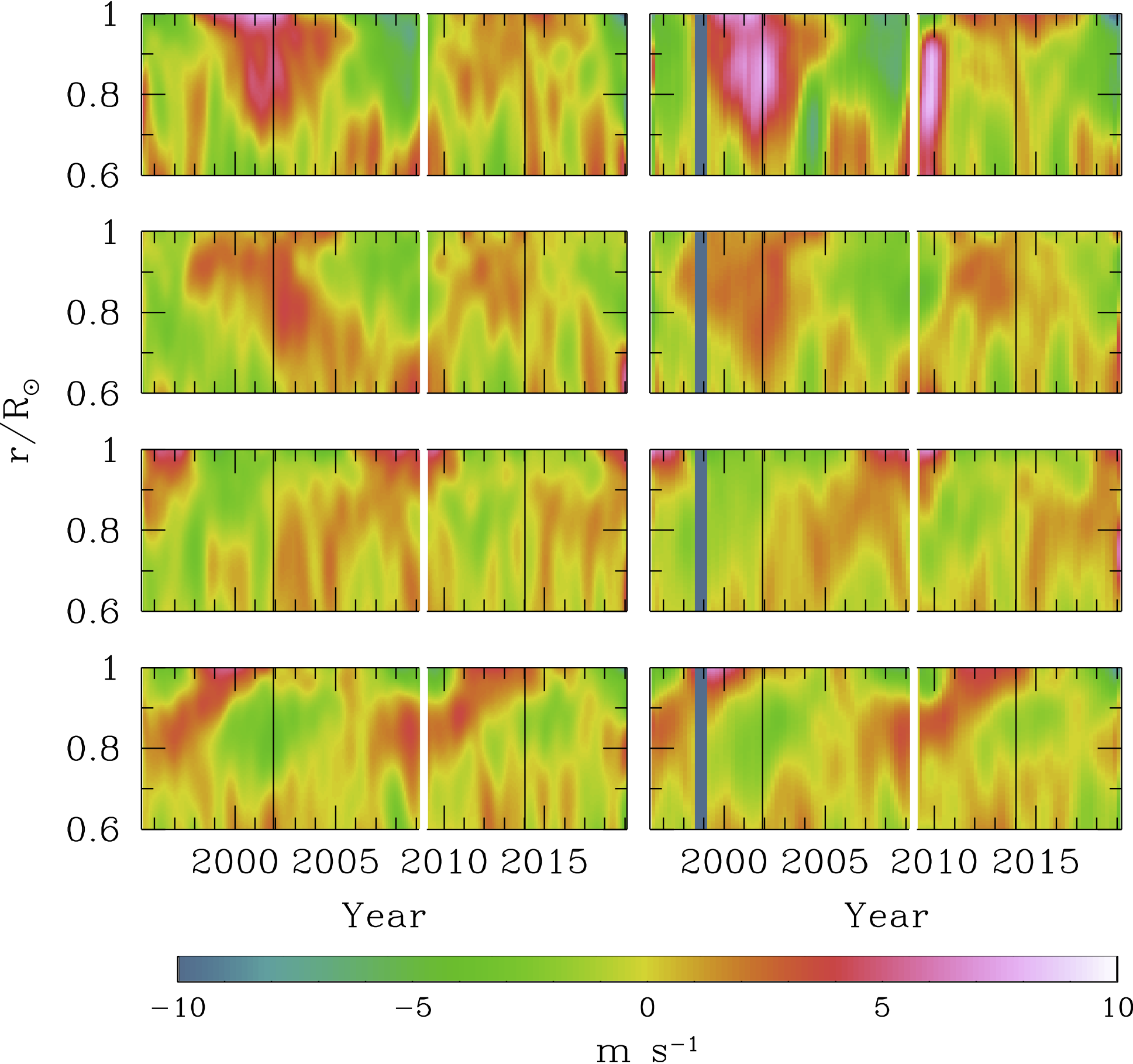}
\caption{The same as Fig.~\ref{fig:zonl} but plotted a a function of time and radius
at, 
from bottom to top, latitudes of $15^\circ$, $30^\circ$, $45^\circ$ and $60^\circ$.
}
\label{fig:zonr}
\end{figure}

The zonal flow velocities at a few different depths as a function of time and
latitude are shown in Fig.~\ref{fig:zonl}. If we look at the topmost
panels, a few features are clear immediately:
despite subtracting a smaller high-latitude average, the poleward flows in
Cycle~24 were much weaker than those in Cycle~23. 
In Cycle~23, the poleward prograde branch end abruptly, and a retrograde band appears
in its place close to the end of the cycle; this change is not seen yet
for Cycle~24, if anything the prograde band appears to have become stronger.
We can also see the
beginnings of what will probably be the low-latitude equator-ward branch of the
zonal flow of Cycle~25. 
However, in Cycle~24  we do not see the smaller ``tuning fork'' type 
equatorward branch at even lower latitudes, that in Cycle~23 was formed around 2005
just as the main equatorward branch became strong. 
Thus the changes in the zonal flows
between Cycle~23 and 24 are not merely quantitative, but they are qualitatively
different as well.  It is possible that the higher average active-latitude
rotation rate in Cycle~24 is hiding this tuning-fork like feature.
Looking deeper, we can see that the
high latitude poleward branch is not well defined at $0.8 R_\odot$.
{The HMI results for $r=0.90R_\odot$ and $0.95R_\odot$ also do not show
a clear poleward branch at high latitudes, but this could be because the
average rotation rate during Cycle 24 is biased by some part of MDI data
during the beginning of Cycle 24, which shows up as a prominent feature at
high latitudes around 2010. In fact, by artificially moving the start of Cycle 24
at 2010.5 to cut out the MDI contribution the poleward branch shows up just like
that in GONG results.}
The  MDI and HMI results show a  
 prominent feature at high latitudes at the beginning of Cycle 24.
This is most likely due to systematic differences between the MDI and HMI data sets.
It should be noted that the first 1.5 years data in Cycle 24 is from MDI and the
rest from HMI. This feature is also seen at $60^\circ$ latitude in Fig.~\ref{fig:zonr}.

In Fig.~\ref{fig:zonr} we show the zonal flow velocities as a function of
radius at different latitudes. As can be seen, the radial pattern of the zonal
flows is qualitatively similar at low latitudes in both cycles, with prograde bands rising
from deep inside the convection zone to the surface as a function of time. 
{At the edge of the active latitudes ($30^\circ$, second row from bottom) a rising
band is seen which starts rising from the convection-zone base just about the
same time as the equatorial prograde flow reaches the solar surface, just before
solar maximum, and straddles the solar minimum.  Thus in a sense, this $30^\circ$
prograde band is precursor of the next cycle. Although this flow is less clear
in Cycle~24, it is still visible showing the beginnings of Cycle~25.}
Above the active latitudes, and we have shown the results at $45^\circ$, the solar
maximum marks the ``sinking'' of the prograde flow. This band appears less
well-defined in Cycle~24 than in Cycle~23.

\begin{figure}[]
\epsscale{0.95}
\plottwo{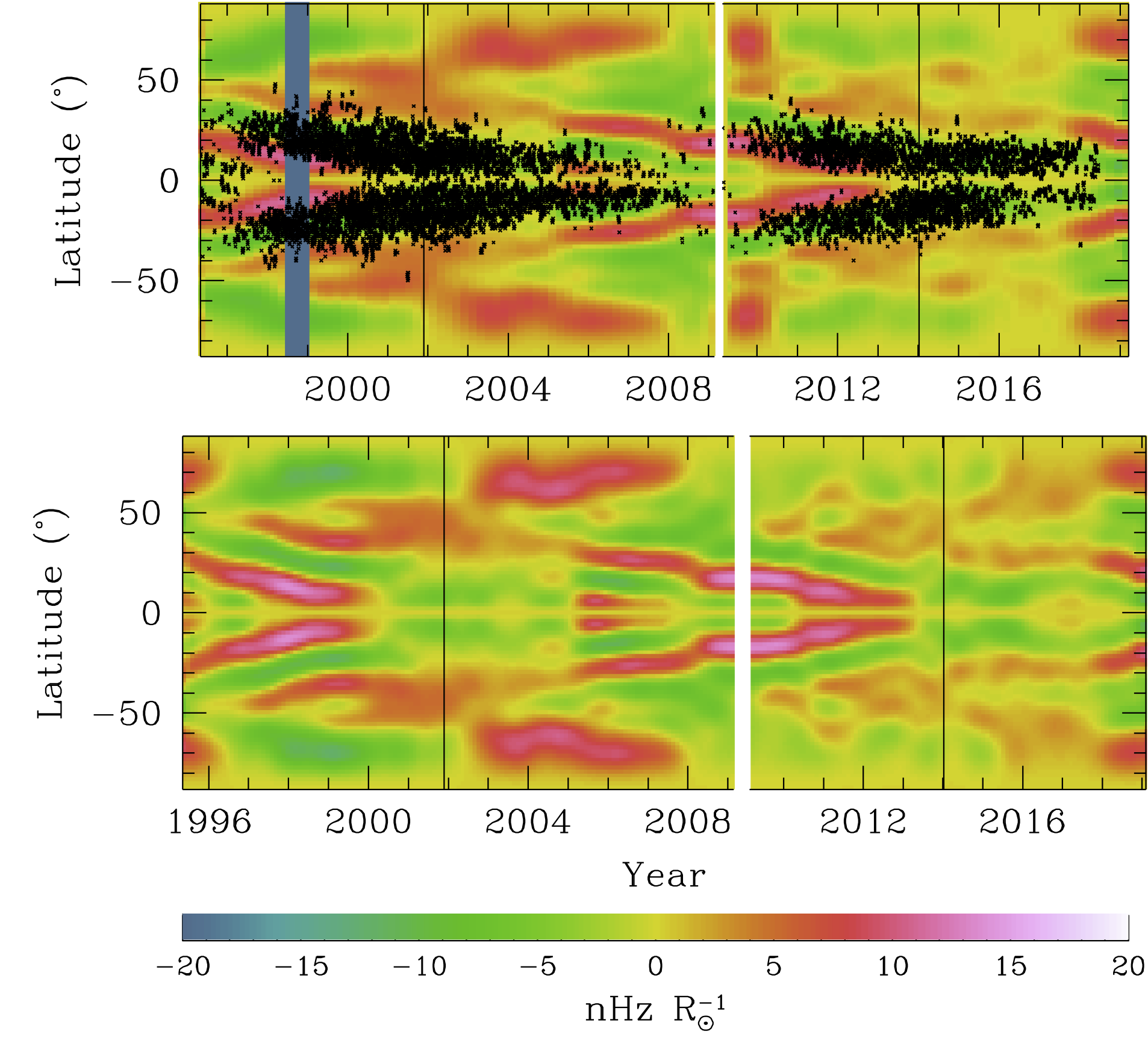}{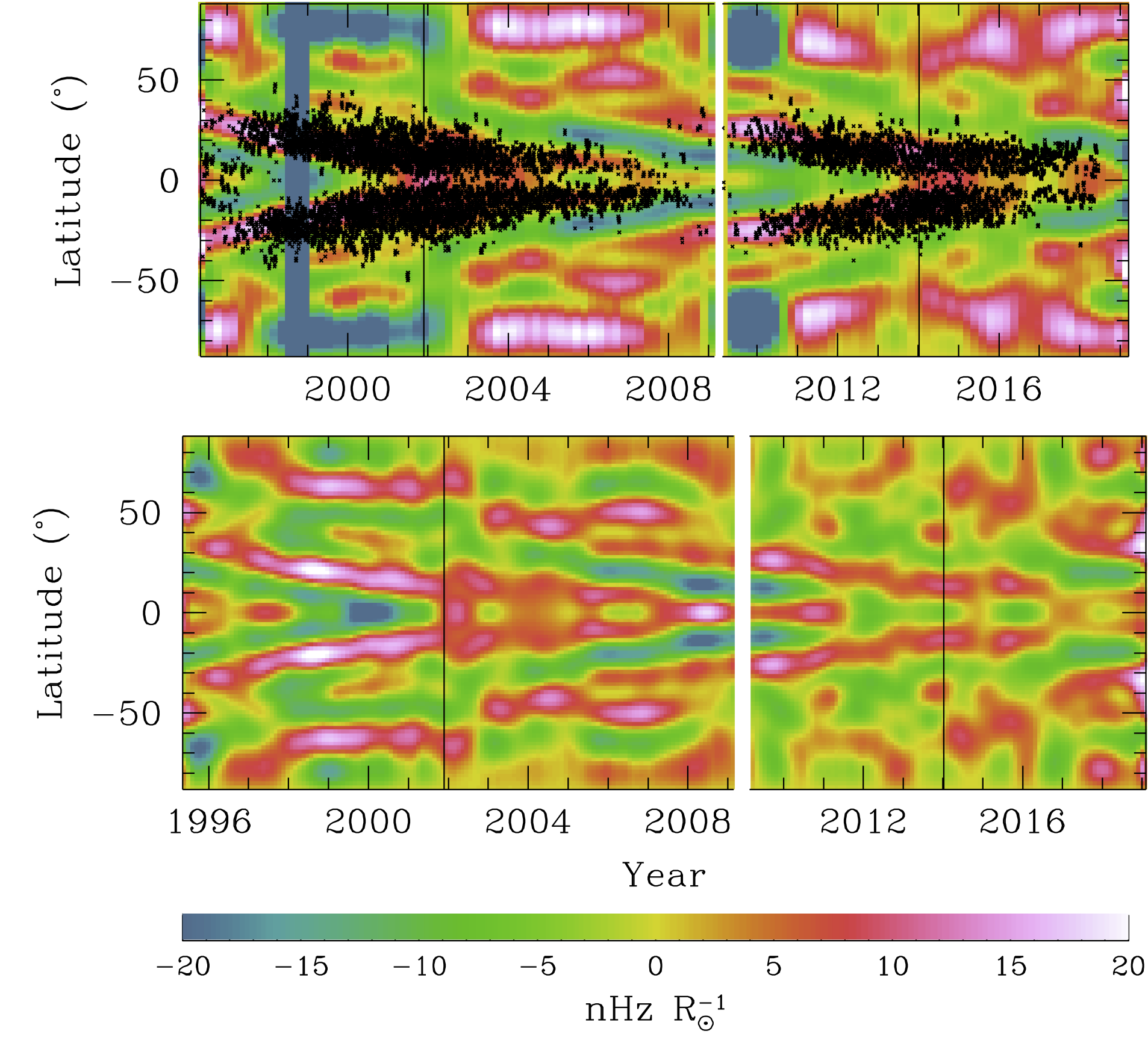}
\caption{{\bf Left column:} The latitudinal gradient, $(1/r)(\partial\delta\Omega/\partial|\theta|)\cos\theta$
	of the zonal flow rate at 0.95\rsun. {\bf Right column:} The radial gradient, $(\partial\delta\Omega/\partial r)\cos\theta$
of the zonal flow rate at 0.95\rsun. In both columns, the upper panels are results from MDI and HMI,
the lower panel shows results from GONG. The position of sunspots are marked in the upper
panel.
}
\label{fig:zoncdt}
\end{figure}

Also of interest are the spatial gradients of the zonal flows, we show those in
Fig.~\ref{fig:zoncdt}. On examining the latitudinal gradient of the zonal flow rate,
it is clear that the gradients were stronger and more well defined in Cycle~23 than
in Cycle~24. This is particularly true for the high-latitude band. The high-latitude band
shows an  abrupt change from positive to negative close to the end of Cycle~23, we do not
see such a change in the Cycle~24 data, most likely indicating that we are not close
to the end of Cycle~24. The minimum after Cycle~23 occurred about one year after the
change of sign of the high-latitude gradient. If that feature is an indicator of a shift
in activity, we are not close to the minimum after Cycle~24. Indeed, \citet{upton} using different 
indicators claim that Cycle~24 will end closer to the end of 2020 or in early 2021.
The radial gradients are more confusing as can be seen in the right-hand panel of
Fig.~\ref{fig:zoncdt}.  The high-latitude results obtained
with MDI and HMI data do not agree with those obtained with GONG data, however, the
results for the active latitudes do agree. Again, Cycle~23 had larger 
gradients (both positive and negative) than Cycle~24.
And as had been seen for Cycle~23 by \citet{hma2008}, sunspots appear where the 
the gradients of the zonal flow are large. A closer examination of both cycles show that
sunspots appear in the low-latitude regions that have a large negative latitudinal gradient
 and a  large positive radial gradient.

\section{The tachocline}
\label{sec:tach}

\subsection{The analysis technique}
\label{subsec:analy}

The tachocline is too thin to be resolved with the usual inversion techniques
\citep{sasha, schou1998, hma1998}, consequently like in earlier works
\citep{hma1998,hmasbtach2011} we use a forward modelling technique
to model the tachocline at each epoch to determine the properties and their time-variations.
This therefore involves adopting a model of the tachocline, determining the
splitting coefficients for different free parameters of the tachocline and
comparing the model coefficients with the observed ones to find the set of
parameters that give the lowest mismatch with the observations, as quantified
by the value of $\chi^2$. We use splittings of all modes with lower turning
points between $0.5R_\odot$ and $0.9R_\odot$. The lower limit is imposed
because of increased uncertainty of the splittings of modes that probe deeper;
the upper limit is set so that the signature of the near-surface shear layer
does not dominate the signal of the splitting coefficients.

We use the method of simulated annealing \citep{anneal1,anneal2} to minimize $\chi^2$.
The algorithm uses randomly generated values of the parameters, and we assume
that the parameters have Gaussian priors with the mean and width of the
Gaussian determined from inversions of rotational splittings. While inversions
do not resolve the tachocline very well, they do define it.  Since
the tachocline has many parameters, there is the likelihood of the solution
being trapped in a local minimum; to avoid this, we make 80 attempts using
different sequences of random numbers in the annealing procedure to find the
global $\chi^2$ minimum.
We determine the uncertainties using a  
traditional boot-strapping method where we simulated many realizations of the
observations, fit each one of them in exactly the same manner as the original
data and use the spread as a measure of uncertainty. 

\subsection{The tachocline model}
\label{subsec:model}

We consider both one-dimensional and two dimensional models for the tachocline.
As in our earlier work \citep{basu1997, hma1998, hmasbtach2011}, we model the tachocline as
\be
\Omega_{\rm tach}=\frac{\Delta\Omega}{1+\exp[(r_d-r)/w]},
\label{eq:tach1}
\ee
where $r$ is radius, $r_d$ the mean position of the tachocline, $\Delta\Omega$ the
jump in rotation rate between radii $r$ where $r < r_d$ and $r > r_d$, and $w$ the
width such that the rotation rate changes from a factor $1/(1+e)$ of the
maximum to a factor $e/(1+e)$ in the range $r=r_d-w$ and $r=r_d+w$. The same
model can be applied in a latitude dependent manner with $\Delta\Omega$, $r_d$
and $w$ made functions of co-latitude \citep{hma1998, hmasbtach2011}.

Another model that has been used in literature \citep{sasha,paulchar} is
\be
\Omega_{\rm tach}={\Delta\Omega}\frac{1}{2}{\left(1+\mbox{erf}\left[\frac{2(r-r_d)}{w}\right]\right)}, 
\label{eq:sasha}
\ee
where $w$, the width is now the radial extent over which 0.84 of the full
transition in the rotation rate takes place, i.e., a change from 0.08 to 0.92
of $\Delta\Omega$, this is different from our model, where the width is defined
as {\it half} the width over which $\Delta\Omega$ changes from 0.269 to 0.731
of its value.  Thus when we fit Eq.~\ref{eq:sasha} to the data, we would expect the value
of $w$ to be different but related to the width that we define in our model.
The two functions have similar shape, but the width in this model is about 5
times that in the model in Eq.~\ref{eq:tach1}.
We have tried this model too, and aside from the scaling of width there is no
significant difference in the fitted parameters.

The tachocline  has both a radial and a latitudinal dependence, in
particular, the change in rotation rate, i.e., the jump,  across the tachocline is strongly
dependent on latitude.
An added complication is that the rotation rate of the
Sun in regions away from the  tachocline can have radial
gradients. We choose a 2D model that is similar to our earlier works \citep{hmasbtach2011}:
\be
\Omega(r, \vartheta)=\begin{cases}{\Omega_c+ \frac{\Delta\Omega}{1+\exp[(r_d-r)/w]}}& \mbox{if } r\le0.7R_\odot\\
\Omega_c +B(r-0.7)
+{\Delta\Omega\over {1+\exp[(r_d-r)/w]}}& \mbox{if }  0.7 < r\le0.95R_\odot\\
\Omega_c+0.25B -C(r-0.95)
+{\Delta\Omega\over {1+\exp[(r_d-r)/w]}}& \mbox{if } r>0.95R_\odot
\end{cases}
\label{eq:tach2d}
\ee
Where,  $\Omega_c$, $B$, $C$ are the three parameters defining the smooth part of
rotation rate while as usual $\Delta\Omega$, $r_d$ and $w$ define the tachocline.
Here $B$ defines the gradient in deep convection zone,
while $C$ is the gradient in the near surface shear layer. Note that the coefficients
$B$ and $C$ are fitted for each epoch and thus can be time dependent.

The co-latitude dependence of the different parameters are:
\begin{align}
\Delta\Omega=&\Delta\Omega_3P_3(\vartheta)+\Delta\Omega_5P_5(\vartheta),\label{eq:delo}\\
r_d=&r_{d1}+r_{d3}P_3(\vartheta),\label{eq:rd}\\
w=&w_1+w_3P_3(\vartheta)\label{eq:w},
\end{align}
where using the definition of splitting coefficients \citep[cf.,][]{ritzwoller1991}
\begin{align}
P_3(\vartheta)&=5\cos^2\vartheta-1,\label{eq:p3}\\
P_5(\vartheta)&=21\cos^4\vartheta-14\cos^2\vartheta+1, \label{eq:p5}
\end{align}
and $\vartheta$ is the co-latitude. The difference between this model and the
earlier ones lie in the definition of $B$, in this model $B$ does not have a
latitudinal variation, while in the earlier work $B$ was defined as $B=
B_1+B_3P_3(\vartheta)+B_5P_5(\vartheta)$. More importantly in our current model,
$\Delta\Omega$ does not have latitude-independent term $\Delta\Omega_1$. These
changes were made because we found that the marginalized probability distribution
functions of these
quantities were rather flat. Omitting these parameters helped stabilize the fits
for the other quantities.
It should be noted that $\Delta\Omega$ represents the variation in $\Omega$ in
region near the tachocline and is not affected by rotation rate near the surface
which would be reflected in the smooth part of $\Omega$
modeled by parameters $B$ and $C$.

\subsection{Results}
\label{subsec:res}

\begin{figure}
\epsscale{0.5}
\plotone{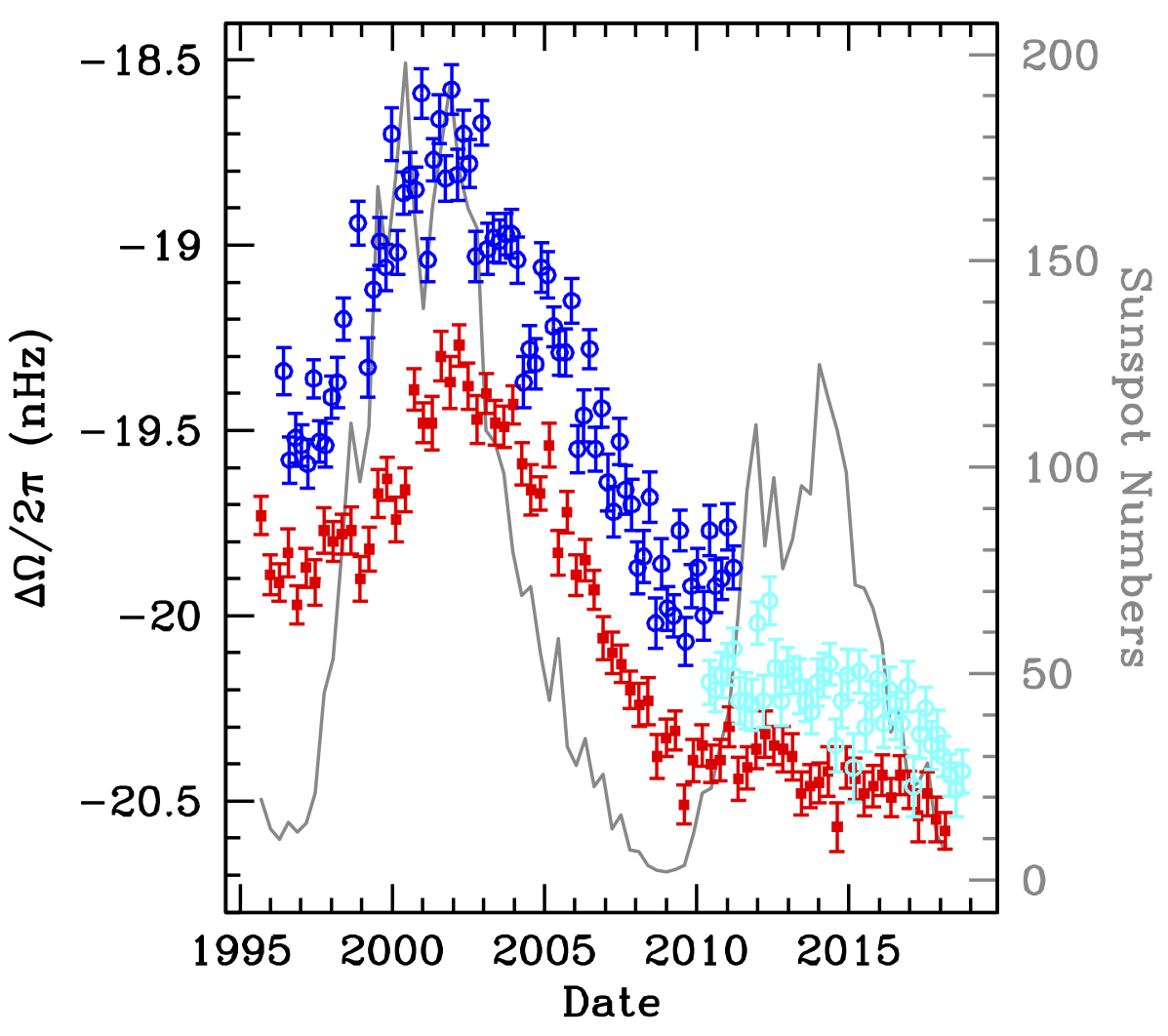}
\caption{The variation of the change in rotation rate across the tachocline as
a function of time as
seen in splitting coefficient $c_3$. Red points are results obtained using GONG data,
deep blue with MDI data and cyan with HMI data. The systematic difference between MDI and GONG
results can be reduced to the HMI-GONG level by restricting the MDI modes to $l < 120$ as
mentioned in Section~\ref{sec:rot}.
The gray curve in the
background is the sunspot number that can be read from the
scale of the ordinate on the right hand side of the box.
}
\label{fig:c3jump}
\end{figure}
\begin{figure}
\epsscale{0.55}
\plotone{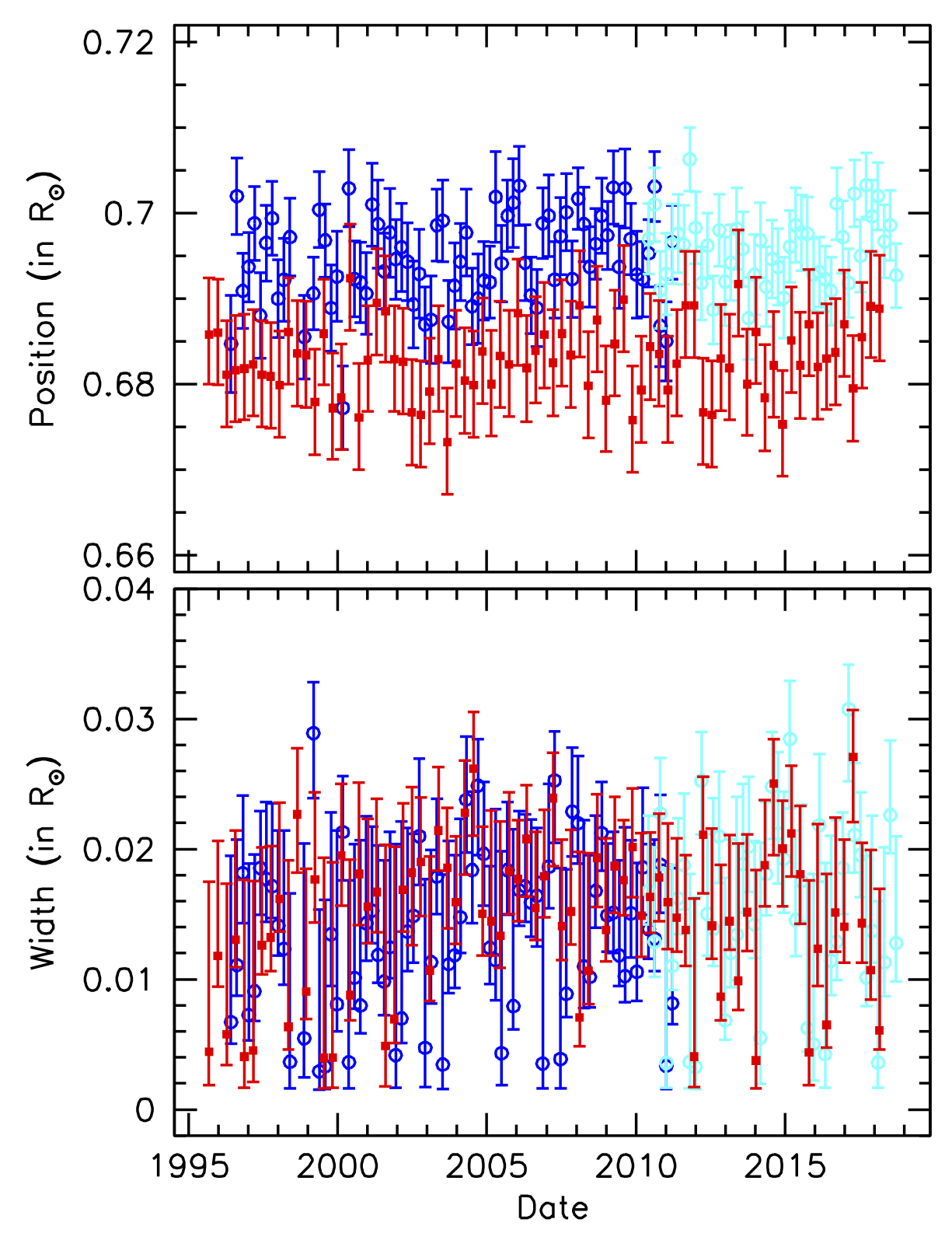}
\caption{The variation of the position and width of the tachocline as
determined from splitting
 coefficient $c_3$. Red points are results obtained using GONG data,
deep blue with MDI data and cyan with HMI data.
}
\label{fig:c3rd}
\end{figure}

In keeping with earlier work, we first fitted the $c_3$ coefficient, which shows the
largest signal of the tachocline, alone to the model in Eq.~\ref{eq:tach1}.
 Although, the $c_3$ results in Fig~\ref{fig:coef}
are for a much shallower radius, the change in $c_3$ with time leads us to expect 
a variation in tachocline properties with time.
The results of $\Delta\Omega$, the jump across
the tachocline are shown in Fig.~\ref{fig:c3jump}. Clearly,
the quantity shows a statistically significant time variation. Also note that 
while in Cycle~23  $\Delta\Omega$ shows a variation that reflects the change
in solar activity indices, the change is much more subtle in Cycle~24.
Also note that although there is a small systematic difference between the GONG
and MDI+HMI results, the time variations remain the same.
In contrast to $\Delta\Omega$, the position and width of the tachocline shows no
discernible change with time, as is shown in Fig.~\ref{fig:c3rd}, and
and in particular, we do not see any
significant ``pulsations'' in the tachocline position as has been hypothesized 
\citep{dejager}.
The results do not change if we use the tachocline model in Eq.~\ref{eq:sasha} instead
of Eq.~\ref{eq:tach1}.

\begin{figure}
\epsscale{0.55}
\plotone{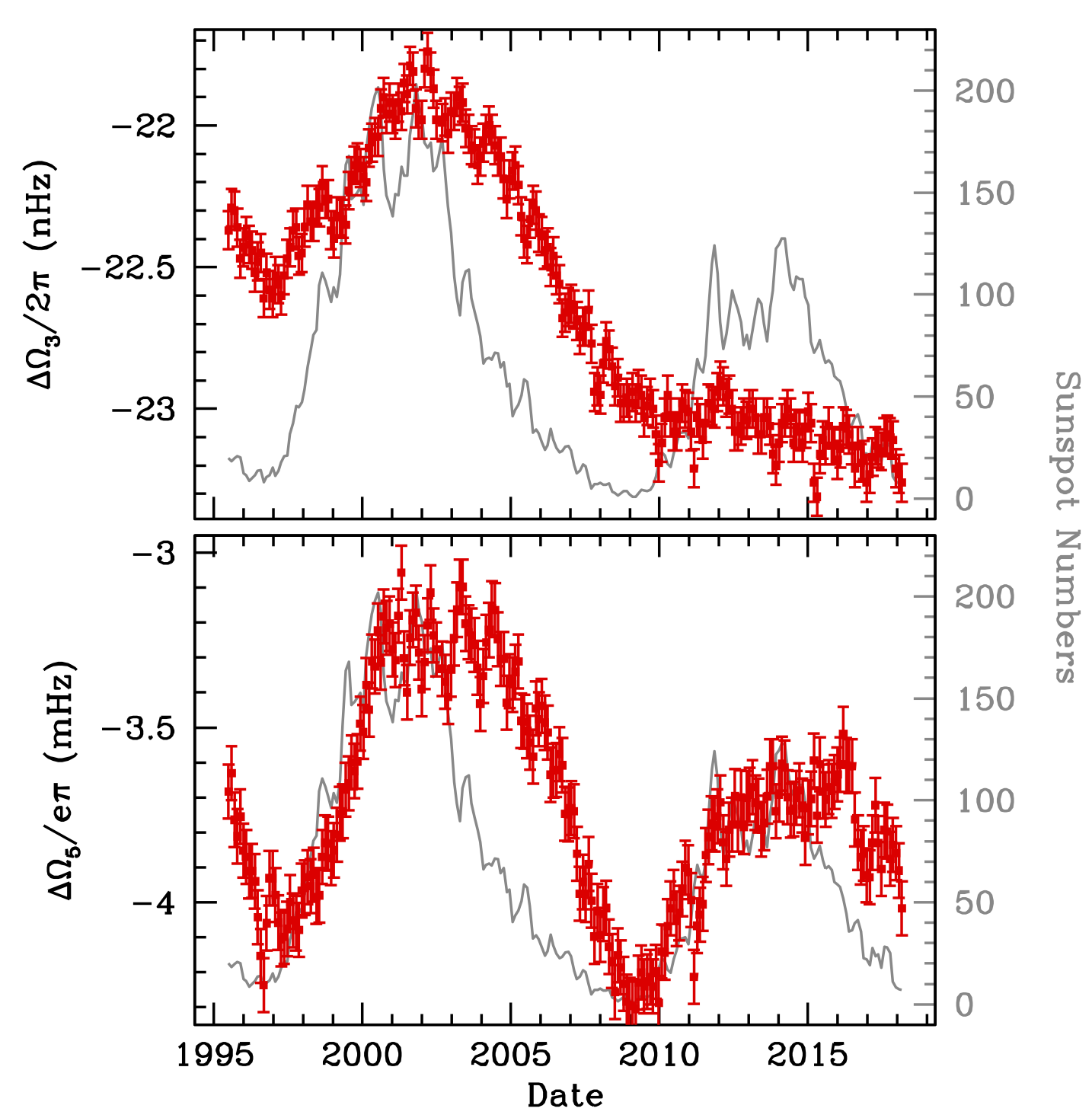}
\caption{The two components of the jump across the tachocline $\Delta\Omega_3$ and
$\Delta\Omega_5$ plotted as a function of time. Only GONG results are shown. The gray 
line in the background is the SSN plotted on the scale to the right of the panels.
}
\label{fig:om35}
\end{figure}

\begin{figure}
\epsscale{1.0}
\plottwo{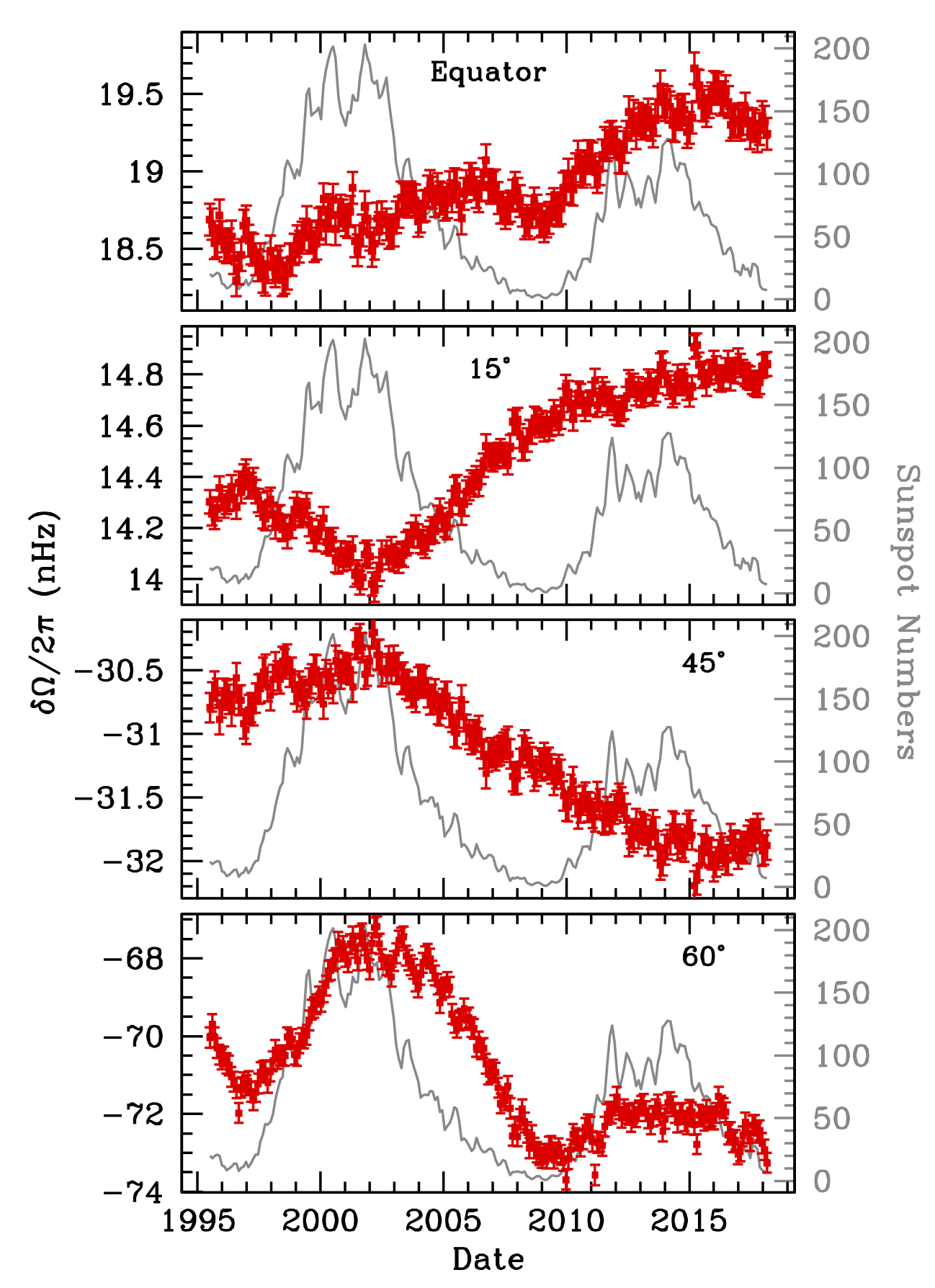}{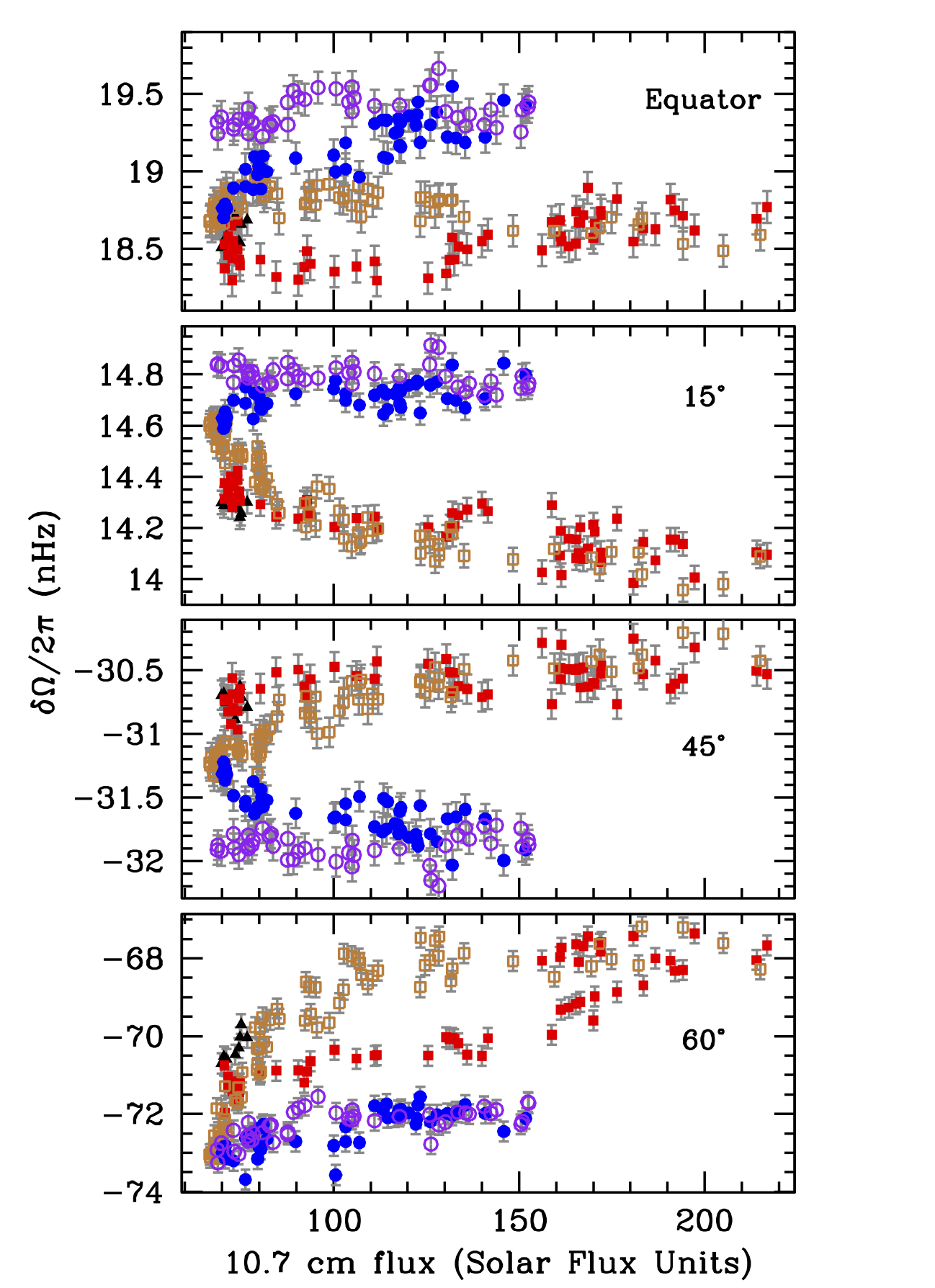}
\caption{ {\bf Left:}
	The change in the tachocline with time at different latitudes {is shown by
	red points.}
Only GONG results have been shown
for clarity. The gray curve in the
background is the SSN  that can be read from the
scale of the ordinate on the right hand side of the panels.
{\bf Right:} The change in the tachocline as a function of 
the flux of 10.7 cm radio emission, which is a proxy for solar activity.
In each panel,
the black triangles mark the descending phase of Cycle~22, red filled squares the ascending phase of
Cycle~23, brown open square show the descending phase of Cycle~23, blue filled circles and
purple open squares are respectively the ascending and descending phases of Cycle~24. Only
GONG results are shown. The flux is in  Solar Flux Units (SFU), with 1~SFU~$= 10^4$ Jansky
$=10^{-22}$ W~m$^{-2}$~Hz$^{-1}$.
}
\label{fig:2djump}
\end{figure}

The $\Delta\Omega_3$ and $\Delta\Omega_5$ parameters obtained by fitting the two-dimensional model  
(i.e., fitting Eq.~\ref{eq:tach2d}) are shown in Fig.~\ref{fig:om35}. We can see that $\Delta\Omega_3$
is basically unchanged from that shown in Fig.~\ref{fig:c3jump} where we fitted a much
simpler model only to $c_3$. This shows that the results are robust. We find that 
 the two components
of the jump across the tachocline follow the time variation of the coefficients shown in 
Fig.~\ref{fig:coef}, i.e., $\Delta\Omega_5$ follows the sunspot number, while $\Delta\Omega_3$ does, not.
Since $\Delta\Omega_3$ and $\Delta\Omega_5$ determine the latitudinal behavior of the
jump, we expect the different latitudes to behave quite differently in the two cycles. This is
indeed the case, as is shown in  Fig.~\ref{fig:2djump}.
We only show the results for $\Delta\Omega$ since like earlier $r_d$ and $w$ do not show any
significant time variation. Note that at no latitude does the time variation of $\Delta\Omega$
during Cycle~24 look like the variation in Cycle~23. If there were a one-to-one relation between
solar activity and the properties of the tachocline, we would have expected Cycle~24 results to
be a repeat of the Cycle~23 results, but with a smaller change between solar maximum and minimum
in keeping with the overall lower activity level of Cycle~24; instead we see that low and intermediate
latitudes, the nature of the change is completely different.
The results are also plotted against
the 10.7 cm flux,
 and it is clear that at the same level of activity,
the jump of the rotation rate across the tachocline was very different. 

\begin{figure}
\epsscale{0.6}
\plotone{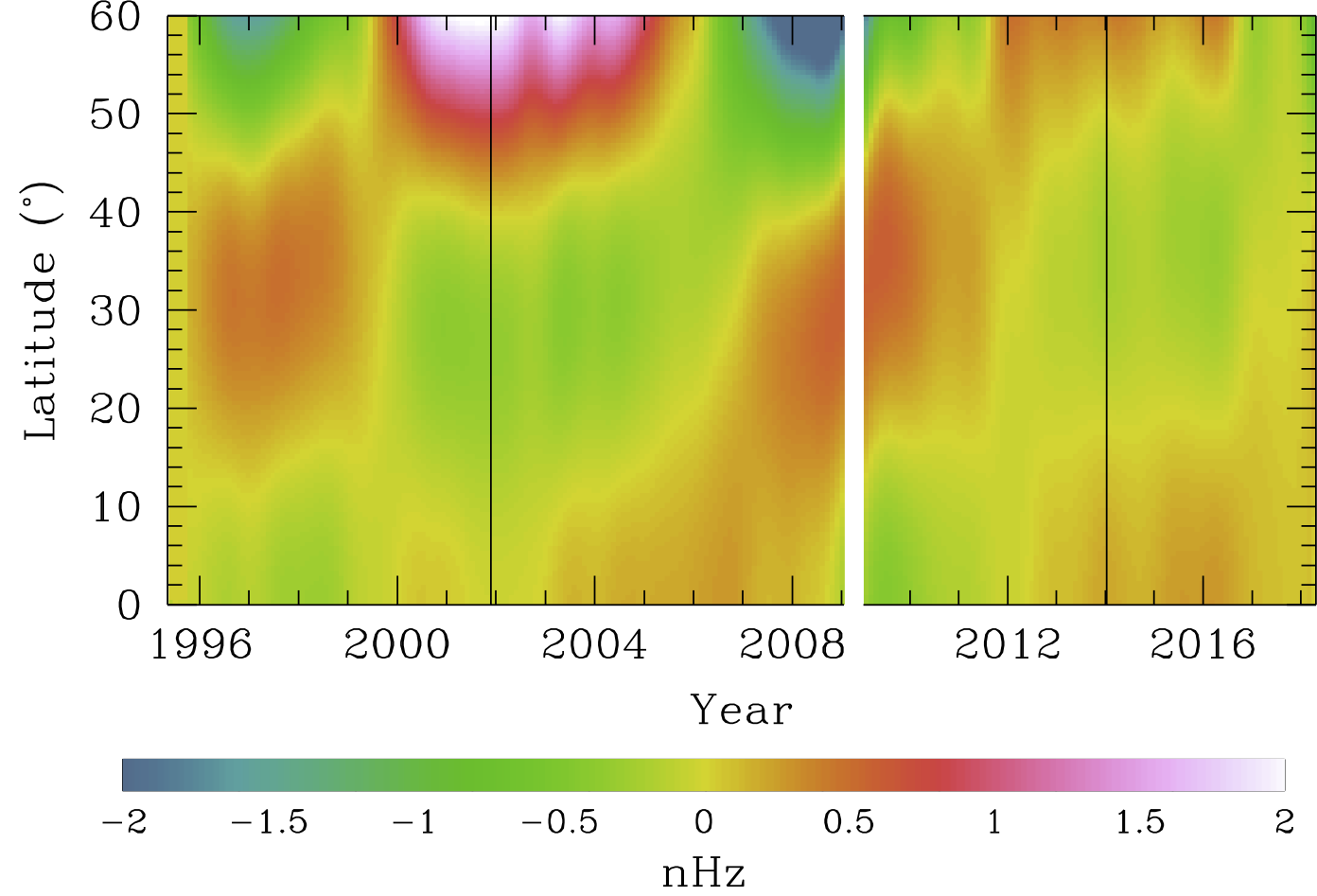}
\caption{The change in the jump across the tachocline plotted as a function
of time and latitude. In order to show the change clearly, the average
value of jump during each solar cycle has been subtracted from the
value at each latitude. The blank space demarcates the two cycles.
This way of plotting
emphasizes the behavior with time, rather than the overall magnitude of the
tachocline, which as shown in Fig.~\ref{fig:2djump}, differed in the two solar cycles.
}
\label{fig:tachzon}
\end{figure}

The change in the rotation rate across the tachocline can actually be plotted the way
we have plotted the zonal flows, {by subtracting out the average jump over
each cycle at each latitude.} This is shown in Fig:~\ref{fig:tachzon}, it should be
noted that unlike zonal flow figures, the results at different latitudes are at
different radii, the radius at each latitude is $r_d$ obtained from the
fits to Eq.~\ref{eq:tach2d}. {Subtracting the average jump over each
cycle shows that the time-variation during each cycle is more similar, however
  a
few features stand out immediately:} (1) there are alternating bands of positive
and negative change, and the behavior in time is latitude dependent.  (2) At
the solar maximum, the behavior of the tachocline above active latitudes is
much more marked in Cycle~23 than that at the maximum of Cycle~24. (3) At the start
of each cycle, there is a
positive band in the mid-latitude regions, that in Cycle~24 reaches to higher latitudes
than in Cycle~23; the band is also somewhat stronger in Cycle~24. (4) The
mid-latitude negative band is stronger in Cycle~23 than in Cycle~24. (5) The two intense
high-latitude negative region in Cycle~23 are barely noticeable in Cycle~24.
{Since  Figure~\ref{fig:tachzon} shows  residuals when the average jump over each cycle is subtracted
out, there is little contribution in Cycle~24 from the splitting
coefficient $c_3$ which is almost constant during this period. As a result,  the contribution
from coefficient $c_5$ dominates. During cycle 23 on the other hand, the major contribution is from
the coefficient $c_3$.
Thus while the qualitative  differences in the nature of the
tachocline between the two cycles remain, though in this form that are  not
as dramatic as the changes shown 
 in Figure~\ref{fig:2djump}.}

\section{Summary and Conclusions}
\label{sec:conc}

Our analysis of solar rotation using helioseismic data has revealed some 
surprises. If we confine our attention to oscillation modes that have lower turning 
points in the outer convection zone, we find that splitting coefficient $c_3$ has behaved very
different from the others; it does not show a monotonic change in activity and instead has 
kept decreasing since the maximum of Cycle~23. Changes in $c_3$ in the 
upcoming Cycle~25 should tell us if $c_3$ varies periodically at all. Coefficients $c_5$ and
$c_7$ show an abrupt change at the beginning of solar Cycles~23 and 24, and  
again Cycle~25 will reveal whether these can be used as a marker of the onset of the
solar cycle.

The average solar rotation rate over Cycle~23 was quite different than the average over the
current span of Cycle~24. Cycle~23 had higher rotation velocities at high latitudes,
but lower at the low latitudes. The difference in the average rotation rate affects the
zonal flow pattern that is seen in the two cycles. As a result, to see the zonal flows
clearly, particularly during Cycle~24 we subtract separate averages over each cycle. 
The fact that the zonal flow pattern and velocities
depend on the average subtracted is an indication of the fact that the way zonal
flows are revealed is not robust and the subtraction of an incompatible average can 
yield misleading results.
Note that some of the differences could be due to the fact that Cycle~24 has not 
ended yet.

The zonal flow pattern in Cycle~24 was much less defined, and had lower velocities
than Cycle~23. In particular, the high-latitude poleward branch is very weak. Our results
confirm those of \citet{rachel2018}. Like \citet{rachel2018} we see the start of the
dynamical effects of solar Cycle~25. The signs are clearer when the flows are examined
as a function of radius at a latitude of about $30^\circ$. In Cycle~23, a prograde 
branch appeared after the maximum at this latitude near the tachocline and that rose up
to form the low-latitude prograde flow in Cycle~24; we see that such a band did form
after the Cycle~24 maximum too. The band is however, quite weak, leading us to believe
that Cycle~25 will be quite weak. 

The spatial gradients of the zonal flows in Cycle~23 and 24 show that sunspots
are formed where and when a large negative latitudinal gradient and a
large positive radial gradient arise at low latitudes in near surface layers.
However, these gradients are different for
the  two cycles. We are yet to see a change in the sign of the latitudinal
gradient of the poleward branch;  Cycle~23 ended about a year after that shift,
and its absence in Cycle~24 so far leads us to believe that Cycle~25 is not
imminent, unlike the prediction of \citet{rachel2018}, but could be delayed as
predicted by \citet{upton}. 

The dynamics of the tachocline were very different in the two solar cycles.
While we did not find any statistically significant changes in the position or
width of the tachocline, we find that the change in rotation rate across the
tachocline had a different variation in Cycle~23 compared with Cycle~24. The
changes cannot be explained as just a dependence on solar activity --- the jump
in rotation, $\Delta\Omega$ is different in the two cycles for the same value
of the sunspot number or the 10.7 cm flux.  This is mainly a result of the time
variation of splitting  coefficient $c_3$, the one with the largest signature
of the tachocline; the 
jump in tachocline resulting from the splitting coefficient
$c_5$ shows similar behavior in the two cycles and basically follows the
the solar activity indices, albeit with a small delay at the descending phase.

\acknowledgments This work utilizes data obtained by the GONG program, managed
by the National Solar Observatory, which is operated by AURA, Inc. under a
cooperative agreement with the National Science Foundation. The data were
acquired by instruments operated by the Big Bear Solar Observatory, High
Altitude Observatory, Learmonth Solar Observatory, Udaipur Solar Observatory,
Instituto de Astrofísica de Canarias, and Cerro Tololo Interamerican
Observatory.  This work also utilizes data from the MDI instrument on board
SOHO and the HMI instrument on SDO.  SOHO is a project of international
cooperation between ESA and NASA. HMI data is courtesy of NASA/SDO and the HMI
science team. We also acknowledge the SILSO World Data Center for sunspot
numbers, the Debrecen Observatory for sunspot positions, and the National
Research Council of Canada for the 10.7 cm radio flux.

\facilities{GONG, SOHO (MDI), SDO (HMI)}


\begin{thebibliography}{}
\expandafter\ifx\csname natexlab\endcsname\relax\def\natexlab#1{#1}\fi
\providecommand{\url}[1]{\href{#1}{#1}}
\providecommand{\dodoi}[1]{doi:~\href{http://doi.org/#1}{\nolinkurl{#1}}}
\providecommand{\doeprint}[1]{\href{http://ascl.net/#1}{\nolinkurl{http://ascl%
.net/#1}}}
\providecommand{\doarXiv}[1]{\href{https://arxiv.org/abs/#1}{\nolinkurl{https:%
//arxiv.org/abs/#1}}}

\bibitem[{{Antia} \& {Basu}(2000)}]{antiabasu2000}
{Antia}, H.~M., \& {Basu}, S. 2000, \apj, 541, 442

\bibitem[{{Antia} \& {Basu}(2001)}]{antiabasu2001}
---. 2001, \apj, 559, L67

\bibitem[{{Antia} \& {Basu}(2004)}]{ab2004}
{Antia}, H.~M., \& {Basu}, S. 2004, in ESA Special Publication, Vol. 559, SOHO
  14 Helio- and Asteroseismology: Towards a Golden Future, ed. D.~{Danesy}, 301

\bibitem[{{Antia} \& {Basu}(2010)}]{antiabasu2010}
---. 2010, \apj, 720, 494

\bibitem[{{Antia} \& {Basu}(2011)}]{hmasbtach2011}
---. 2011, \apjl, 735, L45

\bibitem[{{Antia} \& {Basu}(2013)}]{antiabasu2013}
{Antia}, H.~M., \& {Basu}, S. 2013, in Journal of Physics Conference Series,
  Vol. 440, Journal of Physics Conference Series, 012018

\bibitem[{{Antia} {et~al.}(1998){Antia}, {Basu}, \& {Chitre}}]{hma1998}
{Antia}, H.~M., {Basu}, S., \& {Chitre}, S.~M. 1998, \mnras, 298, 543

\bibitem[{{Antia} {et~al.}(2008){Antia}, {Basu}, \& {Chitre}}]{hma2008}
---. 2008, \apj, 681, 680

\bibitem[{{Baranyi} {et~al.}(2016){Baranyi}, {Gy{\H o}ri}, \&
  {Ludm{\'a}ny}}]{baranyi1}
{Baranyi}, T., {Gy{\H o}ri}, L., \& {Ludm{\'a}ny}, A. 2016, \solphys, 291, 3081

\bibitem[{{Basu}(1997)}]{basu1997}
{Basu}, S. 1997, \mnras, 288, 572

\bibitem[{Basu {et~al.}(2012)Basu, Broomhall, Chaplin, \&
  Elsworth}]{basuetal2012}
Basu, S., Broomhall, A.-M., Chaplin, W.~J., \& Elsworth, Y. 2012, Astrophys.
  J., 758, 43

\bibitem[{{Charbonneau} {et~al.}(1999){Charbonneau}, {Christensen-Dalsgaard},
  {Henning}, {Larsen}, {Schou}, {Thompson}, \& {Tomczyk}}]{paulchar}
{Charbonneau}, P., {Christensen-Dalsgaard}, J., {Henning}, R., {et~al.} 1999,
  \apj, 527, 445

\bibitem[{{de Jager} {et~al.}(2016){de Jager}, {Akasofu}, {Duhau},
  {Livingston}, {Nieuwenhuijzen}, \& {Potgieter}}]{dejager}
{de Jager}, C., {Akasofu}, S.-I., {Duhau}, S., {et~al.} 2016, \ssr, 201, 109

\bibitem[{{Gilman}(2005)}]{gilman}
{Gilman}, P.~A. 2005, Astronomische Nachrichten, 326, 208

\bibitem[{{Gy{\H o}ri} {et~al.}(2017){Gy{\H o}ri}, {Ludm{\'a}ny}, \&
  {Baranyi}}]{baranyi2}
{Gy{\H o}ri}, L., {Ludm{\'a}ny}, A., \& {Baranyi}, T. 2017, \mnras, 465, 1259

\bibitem[{{Hill} {et~al.}(2011){Hill}, {Howe}, {Komm}, {Hern{\'a}ndez},
  {Kholikov}, \& {Leibacher}}]{hill2011}
{Hill}, F., {Howe}, R., {Komm}, R., {et~al.} 2011, in IAU Symposium, Vol. 271,
  Astrophysical Dynamics: From Stars to Galaxies, ed. N.~H. {Brummell}, A.~S.
  {Brun}, M.~S. {Miesch}, \& Y.~{Ponty}, 15--22

\bibitem[{{Hill} {et~al.}(1996){Hill}, {Stark}, {Stebbins}, {Anderson},
  {Antia}, {Brown}, {Duvall}, {Haber}, {Harvey}, {Hathaway}, {Howe}, {Hubbard},
  {Jones}, {Kennedy}, {Korzennik}, {Kosovichev}, {Leibacher}, {Libbrecht},
  {Pintar}, {Rhodes}, {Schou}, {Thompson}, {Tomczyk}, {Toner}, {Toussaint}, \&
  {Williams}}]{hill1996}
{Hill}, F., {Stark}, P.~B., {Stebbins}, R.~T., {et~al.} 1996, Science, 272,
  1292

\bibitem[{{Howard} \& {Labonte}(1980)}]{howard1980}
{Howard}, R., \& {Labonte}, B.~J. 1980, \apjl, 239, L33

\bibitem[{{Howe} {et~al.}(2013{\natexlab{a}}){Howe}, {Christensen-Dalsgaard},
  {Hill}, {Komm}, {Larson}, {Rempel}, {Schou}, \& {Thompson}}]{howe2013}
{Howe}, R., {Christensen-Dalsgaard}, J., {Hill}, F., {et~al.}
  2013{\natexlab{a}}, \apj, 767, L20

\bibitem[{{Howe} {et~al.}(2013{\natexlab{b}}){Howe}, {Christensen-Dalsgaard},
  {Hill}, {Komm}, {Larson}, {Schou}, \& {Thompson}}]{howeetal2013}
{Howe}, R., {Christensen-Dalsgaard}, J., {Hill}, F., {et~al.}
  2013{\natexlab{b}}, in Astronomical Society of the Pacific Conference Series,
  Vol. 478, Fifty Years of Seismology of the Sun and Stars, ed. K.~{Jain},
  S.~C. {Tripathy}, F.~{Hill}, J.~W. {Leibacher}, \& A.~A. {Pevtsov}, 303

\bibitem[{{Howe} {et~al.}(2000){Howe}, {Christensen-Dalsgaard}, {Hill}, {Komm},
  {Larsen}, {Schou}, {Thompson}, \& {Toomre}}]{rachel2000}
---. 2000, \apj, 533, L163

\bibitem[{{Howe} {et~al.}(2017){Howe}, {Davies}, {Chaplin}, {Elsworth}, {Basu},
  {Hale}, {Ball}, \& {Komm}}]{howeetal2017}
{Howe}, R., {Davies}, G.~R., {Chaplin}, W.~J., {et~al.} 2017, \mnras, 470, 1935

\bibitem[{{Howe} {et~al.}(2018){Howe}, {Hill}, {Komm}, {Chaplin}, {Elsworth},
  {Davies}, {Schou}, \& {Thompson}}]{rachel2018}
{Howe}, R., {Hill}, F., {Komm}, R., {et~al.} 2018, \apj, 862, L5

\bibitem[{{Komm} {et~al.}(2014){Komm}, {Howe}, {Gonz{\'a}lez Hern{\'a}ndez}, \&
  {Hill}}]{rudi2014}
{Komm}, R., {Howe}, R., {Gonz{\'a}lez Hern{\'a}ndez}, I., \& {Hill}, F. 2014,
  \solphys, 289, 3435

\bibitem[{{Kosovichev}(1996)}]{sasha}
{Kosovichev}, A.~G. 1996, \apjl, 469, L61

\bibitem[{{Kosovichev} \& {Pipin}(2019)}]{sasha2019}
{Kosovichev}, A.~G., \& {Pipin}, V.~V. 2019, \apj, 871, L20

\bibitem[{{Kosovichev} \& {Schou}(1997)}]{agkjs1997}
{Kosovichev}, A.~G., \& {Schou}, J. 1997, \apjl, 482, L207

\bibitem[{{Larson} \& {Schou}(2015)}]{larson}
{Larson}, T.~P., \& {Schou}, J. 2015, \solphys, 290, 3221

\bibitem[{{Larson} \& {Schou}(2018)}]{larson2018}
---. 2018, \solphys, 293, 29

\bibitem[{{Press} {et~al.}(1992){Press}, {Teukolsky}, {Vetterling}, \&
  {Flannery}}]{anneal2}
{Press}, W.~H., {Teukolsky}, S.~A., {Vetterling}, W.~T., \& {Flannery}, B.~P.
  1992, {Numerical recipes in FORTRAN. The art of scientific computing}
  (Cambridge: University Press, |c1992, 2nd ed.)

\bibitem[{{Ritzwoller} \& {Lavely}(1991)}]{ritzwoller1991}
{Ritzwoller}, M.~H., \& {Lavely}, E.~M. 1991, \apj, 369, 557

\bibitem[{{Scherrer} {et~al.}(1995){Scherrer}, {Bogart}, {Bush}, {Hoeksema},
  {Kosovichev}, {Schou}, {Rosenberg}, {Springer}, {Tarbell}, {Title},
  {Wolfson}, {Zayer}, \& {MDI Engineering Team}}]{mdi}
{Scherrer}, P.~H., {Bogart}, R.~S., {Bush}, R.~I., {et~al.} 1995, \solphys,
  162, 129

\bibitem[{{Scherrer} {et~al.}(2012){Scherrer}, {Schou}, {Bush}, {Kosovichev},
  {Bogart}, {Hoeksema}, {Liu}, {Duvall}, {Zhao}, {Title}, {Schrijver},
  {Tarbell}, \& {Tomczyk}}]{hmi}
{Scherrer}, P.~H., {Schou}, J., {Bush}, R.~I., {et~al.} 2012, \solphys, 275,
  207

\bibitem[{{Schou}(1999)}]{schou1999}
{Schou}, J. 1999, \apjl, 523, L181

\bibitem[{{Schou} {et~al.}(1998){Schou}, {Antia}, {Basu}, {Bogart}, {Bush},
  {Chitre}, {Christensen-Dalsgaard}, {Di Mauro}, {Dziembowski}, {Eff-Darwich},
  {Gough}, {Haber}, {Hoeksema}, {Howe}, {Korzennik}, {Kosovichev}, {Larsen},
  {Pijpers}, {Scherrer}, {Sekii}, {Tarbell}, {Title}, {Thompson}, \&
  {Toomre}}]{schou1998}
{Schou}, J., {Antia}, H.~M., {Basu}, S., {et~al.} 1998, \apj, 505, 390

\bibitem[{{Schou} {et~al.}(2002){Schou}, {Howe}, {Basu},
  {Christensen-Dalsgaard}, {Corbard}, {Hill}, {Komm}, {Larsen},
  {Rabello-Soares}, \& {Thompson}}]{schou2002}
{Schou}, J., {Howe}, R., {Basu}, S., {et~al.} 2002, \apj, 567, 1234

\bibitem[{{SILSO World Data Center}(1995--2019)}]{sidc}
{SILSO World Data Center}. 1995--2019, International Sunspot Number Monthly
  Bulletin and online catalogue

\bibitem[{Tapping(2013)}]{tapping2}
Tapping, K.~F. 2013, Space Weather, 11, 394

\bibitem[{{Tapping} \& {Morton}(2013)}]{tapping}
{Tapping}, K.~F., \& {Morton}, D.~C. 2013, in Journal of Physics Conference
  Series, Vol. 440, Journal of Physics Conference Series, 012039

\bibitem[{{Ulrich}(2001)}]{ulrich2001}
{Ulrich}, R.~K. 2001, \apj, 560, 466

\bibitem[{{Upton} \& {Hathaway}(2018)}]{upton}
{Upton}, L.~A., \& {Hathaway}, D.~H. 2018, \grl, 45, 8091

\bibitem[{{Vanderbilt} \& {Louie}(1984)}]{anneal1}
{Vanderbilt}, D., \& {Louie}, S.~G. 1984, Journal of Computational Physics, 56,
  259

\bibitem[{{Vorontsov} {et~al.}(2002){Vorontsov}, {Christensen-Dalsgaard},
  {Schou}, {Strakhov}, \& {Thompson}}]{vorontsov2002}
{Vorontsov}, S.~V., {Christensen-Dalsgaard}, J., {Schou}, J., {Strakhov},
  V.~N., \& {Thompson}, M.~J. 2002, Science, 296, 101

\end{thebibliography}

\end{document}